\documentclass[12pt]{article}
\usepackage{amsmath}
\usepackage{amsthm}
\usepackage{amssymb}
\usepackage{mathrsfs}
\begin{document}
\def\Tm{T^m}
\def\tm{t_m}
\def\Tdm{T^m_d}
\def\tdm{t_m^d}
\def\ot{\otimes}
\def\br{\mathbb{R}}
\def\al{\alpha}
\def\bt{\beta}
\def\th{\theta}
\def\ga{\gamma}
\def\vth{\vartheta}
\def\de{\delta}
\def\lm{\lambda}
\def\b{\beta}
 \def\tp{{\rm tg}({\phi\over 2})}
\def\k{\kappa}
 \def\pp{{\pi\over 2}}
\def\om{\omega}
\def\si{\sigma}
\def\w{\wedge}
\def\od{\sqrt{2}}
\def\Tn{T^n}
\def\tn{t_n}
\def\Tdn{T^n_d}
\def\tdn{t_n^d}
\def\e{\varepsilon}
\def\ti{\tilde}
\def\js{{1\over 4}}
 \def\D{{\cal D}}
\def\I{{\cal I}}
\def\L{{\cal L}}
\def\ri{{\mathrm{i}}}      
\def\S{{\cal S}}
 \def\H{{\cal H}}
\def\G{{\cal G}}
\def\bz{{\bar z}}
\def\E{{\cal E}}
\def\B{{\cal B}}
\def\M{{\cal M}}
 \def\A{{\cal A}}
\def\K{{\cal K}}
\def\J{{\cal J}}
\def\ub{\Upsilon(b)}
\def\tJ{\ti{\cal J}}
\def\stc{\stackrel{.}{+}}
\def\R{{\cal R}}
\def\d{\partial}
\def\la{\langle}
\def\ra{\rangle}
\def\bc{{\mathbb C}}
\def\st{\stackrel{\w}{,}}
\def\lta{\leftarrow}
\def\rta{\rightarrow}
\def\scu{$SL(2,\bc)/SU(2)$~ $WZW$  }
\def\xpm{\partial_{\pm}}
\def\xp{\partial_+}
 \def\xm{\partial_-}
 \def\ps{\partial_{\sigma}}
 \def\1{{\mbox{\boldmath $1$}}}   
  \def\pt{\partial_{\tau}}
\def\be{\begin{equation}}
\def\ee{\end{equation}}
\def\jp{\frac{1}{ 2}}
\def\noi{\noindent}
\def\nl{\nabla}
\def\sve{\sqrt{\vert\e\vert}}
\def\Ad{{\rm Ad}}
 
\def\tD{\Delta^*}
\def\slc{SL(2,{\bf C})}

\begin{titlepage}
\begin{flushright}
{}~
  
\end{flushright}

\vspace{1cm}
\begin{center}
{\Large \bf   $\eta$ and $\lambda$ deformations as $\E$-models}\\ 
[50pt]{\small
{ \bf Ctirad Klim\v{c}\'{\i}k}
\\
Aix Marseille Universit\'e, CNRS, Centrale Marseille\\ I2M, UMR 7373\\ 13453 Marseille, France}
\end{center}

\vspace{0.5 cm}
\centerline{\bf Abstract}
\vspace{0.5 cm}
\noindent   We show that the  so called $\lm$ deformed $\si$-model  as well as the  $\eta$  deformed  one  belong to a class of the  $\E$-models  introduced  in the context of the Poisson-Lie-T-duality. The $\lm$ and $\eta$  theories differ solely by the choice of the Drinfeld double;   for the $\lambda$ model the double is the direct product $G\times G$ while for the $\eta$ model
it is the complexified group $G^\bc$.  As a consequence of this picture, we prove for any $G$ that the target space geometries of the $\lm$-model and of the Poisson-Lie T-dual of the $\eta$-model are related
by a simple analytic continuation.

\vspace{2cm}


\vspace{1cm}


\end{titlepage}
\section{Summary } Consider the actions $S_\eta(g)$ and $S_\lm(g)$ of the so called $\eta$ and $\lm$ deformed $\si$-models on the target of a simple compact Lie group $G$ :
\be S_\eta(g)= \jp\int  d\xi^+d\xi^-(g^{-1} \partial_+ g,(1-\eta R)^{-1}g^{-1}\partial_-g) ,\label{0}\ee
\be S_\lm(g)= S_{WZW}(g) +\lm\int d\xi^+d\xi^- ((1-\lm{\rm Ad}_{g})^{-1}\d_+gg^{-1},g^{-1}\d_-g).\label{1}\ee
Here $g(\xi^+,\xi^-)\in G$,  the derivatives $\d_\pm$ are taken with respect to the light-cone variables $\xi^\pm$, $(.,.)$ is the  Killing-Cartan form on the Lie algebra $\G^\bc$ of $G^\bc$, $R:\G\to\G$ is  the so called Yang-Baxter operator and $S_{WZW}(g)$ is the standard WZW action  
\be S_{WZW}(g):= \jp\int  d\xi^+d\xi^-(g^{-1} \partial_+ g,g^{-1}\partial_-g)+ \frac{1}{12}\int d^{-1}(dgg^{-1},[dgg^{-1},dgg^{-1}]).\ee
The models \eqref{0} and \eqref{1} were respectively introduced in \cite{K02,K09} and \cite{S14}, with the parameters  $\eta$, $\lm$  real and $\vert\lm\vert<1$.

It may seem that the  expression \eqref{1} defines  a $\sigma$-model also on the complexified group $G^\bc$, however,  this is a false appearance. The reason is that   the action $S_\lm$ evaluated on  $G^\bc$-valued configurations takes generically complex values.
However, if we evaluate  $S_\lm$ exclusively  on configurations $p$ with values in the space $P$ of positive definite Hermitian elements of  $G^\bc$   and we take $\lm$  to be a complex number of modulus $1$ then   $-\ri S_{\lm}(p)$  is always real and   defines some  $\sigma$-model on $P$.  Our {\bf Result 1} (the principal one) then states:

\medskip

 {\it The $\si$-model $-\ri S_\lm(p)$  on $P$ for $\lm =\frac{1-\ri \eta}{1+\ri \eta}$  is  the Poisson-Lie T-dual of the $\eta$-model.}

\medskip
 
\noindent  Few remarks are in order:

 {\small  \smallskip \noindent 1)  The replacing  of the unitary argument $g$ by the positive definite  Hermitian one $p$ in \eqref{1}  can be interpreted as a simple analytic continuation of the coordinates parametrizing the Cartan torus; our result therefore  generalizes to any $G$ the $SU(2)$ result of Refs. \cite{HT},\cite{SST} stating that the $\lm$-model  is related by  analytical continuation to the Poisson-Lie T-dual of 
the $\eta$-model.  

\smallskip \noindent 2) It is very probable that our purely bosonic result can be generalized to the supergroup context. This would mean that, up to the  analytic continuation, the $\lm$-deformed $(AdS_5\times S^5)_\lm$ superstring of Ref. \cite{HMS} is the Poisson-Lie T-dual of
 the $\eta$-deformed $(AdS_5\times S^5)_\eta$ superstring  of Ref.\cite{DMV13b}.

\smallskip \noindent 3) For the group $G=SU(N)$, $P$ coincides with the spaces of positive definite Hermitian $N\times N$ matrices.}
 
 \medskip
 
The results of \cite{HT} and \cite{SST}  on the analytic continuation were obtained by  working in appropriate coordinates on the group $SU(2)$ and on its dual Borel group. It  appears extremely difficult to   generalize that method
to  higher dimensional groups because  the action of the dual $\eta$-model (in its  version known before the present paper; cf. Eq. (40) of \cite{K02}) becomes prohibitively complicated in any coordinate system.  To move forward we have to find a completely coordinate-free framework to work with and this turns out to be possible thanks to our following {\bf Result 2}:

\medskip

 {\it The  $\lm$-model  on any simple Lie group target $G$  belongs to the class of the  $\E$-models considered in \cite{KS96,KS97} in the context of the  Poisson-Lie T-duality.}

\medskip
 
 \noindent The next {\bf Result 3} is the consequence of the previous one
   
   \medskip
 
{\it For every simple compact Lie group $G$ there exists a manifold $P$, a distinguished function $H$ on $P$ and two  compatible Poisson structures $\{.,.\}_0$, $\{.,.\}_1$ on $P$  such that the dynamical system  with the phase space $P$, the Hamiltonian $H$ and the Poisson structure 
$\{.,.\}_0+\e\{.,.\}_1$ can be identified with
 
i)	   the principal chiral model on $G$, for $\e=0$;

ii)	  the   $\lm$-model on $G$, for $\e>0$, where $\lm =(1-\e^{\jp})(1+\e^{\jp})$;

iii)	  the   $\eta$-model on $G$, for $\e<0$, where $\eta=(-\e)^{\jp}$.}

 \medskip
 
\noindent We finish by two more remarks:

\medskip
\noindent 4) The statement of Result 3 could be in principle reconstructed  by composing together several facts established already in \cite{DMV13,S14,V15}, however, we  
consider  as an independent result the way how we obtain it directly and naturally from the formalism of the $\E$-models.  

\medskip

\noindent 5) The Poisson structure $\{.,.\}_0+\e\{.,.\}_1$  is the (symplectic version of) the current algebra built on a one-parameter family $\D_\e$ of the Drinfeld doubles of the Lie algebra $\G\equiv$ Lie$(G)$.  The Hamiltonian $H$ is a quadratic expression in the currents and it is {\it completely determined} by the Hamiltonian of the principal chiral model because it does not depend on $\e$.

\section{Introduction}
 
A problem how to deform an integrable non-linear $\sigma$-model on group manifold in a way preserving the integrability  was formulated some forty years ago and it turned out to be a difficult one. Several integrable deformations of the principal chiral model have been found   in the eighties and the nineties  for the simplest case of the group $SU(2)$  \cite{ BFHP,Ch,F,FOZ} but  for long decades no examples were constructed for higher dimensional groups.  Some  effort (see e.g. \cite{Moh}) has been made to determine a complete system  of conditions which a target geometry on a general Lie group must fulfill  in order to guarantee integrability, however,  attempts to find solutions of this complicated  highly overdetermined system of conditions  essentially failed for other groups than $SU(2)$. This situation lasted until 2008  when, in \cite{K09},  the present author  established the integrability of the so-called $\eta$-deformed (or, equivalently, Yang-Baxter) $\sigma$-model  \cite{K02} for any simple compact Lie group target $G$.

 The integrable $\eta$-deformation of the principal chiral model described in \cite{K09}  was generalised  
 to the context of  integrable coset and  supercoset targets in \cite{DMV13} and \cite{DMV13b}, respectively. In particular, the result \cite{DMV13b} has triggered an important activity
 in the field because of its relevance in the AdS/CFT story \cite{ABF,ADT,AM,BBP,CMY,ER,HRT,ISST,KMY14,KMY,LRT,MY,S,vT}. In a short period of few years, several new integrable deformations of the integrable nonlinear $\sigma$-models were obtained, some of them multi-parametric \cite{BKL,DMV15,H15,KY,K14,L,S14}.  In the present paper, we shall concentrate mainly on the integrable deformation of the WZW model proposed in \cite{S14}. 
It is now called the "$\lambda$-deformation", it belongs to a class of $\sigma$-models introduced in \cite{T} and, similarly as in the $\eta$ case,   it was later generalized to the
 integrable  supercoset targets \cite{HMS}.  
 
 Three papers \cite{V15}, \cite{HT} and \cite{SST} have  recently  discussed  the issue of  possible structural relations  between the integrable $\eta$- and $\lambda$-deformations
 and all of them emphasized   the relevance of the concept of the Poisson-Lie T-duality \cite{KS95,KS96} in this context. In particular,  Vicedo  \cite{V15}  studied extensively the case of the $\lm$-model on a non-compact   simple Lie group admitting  the so-called  split Yang-Baxter operator on its Lie algebra and pointed out the existence
 of  the Poisson-Lie T-dual theory\footnote{A second-order action of  this dual  theory is not explicitly given in \cite{V15} because of the  problems with the factorizability of the underlying Drinfeld double.  In this respect, the formula \eqref{emo} of the present paper includes also the case of the non-factorizable doubles and its usefulness for the further development 
 of the results of \cite{V15} looks very probable.}  resembling the  variant of the $\eta$-model  with real poles of the so-called twist functions (the poles of the twist function of the original $\eta$-model \cite{K02,K09} are complex conjugated).  
 On the other hand, Hoare and Tseytlin \cite{HT}Ê and Sfetsos, Siampos and Thompson  \cite{SST} have sticked to the compact  case and showed   that the $\lambda$-deformation on the $SU(2)$ target is related  by an appropriate analytic continuation to the Poisson-Lie T-dual of the $\eta$-deformation. The principal goal of the present paper is to generalize this result of $\cite{HT,SST}$ to any target $G$.

  The other goal of the
 present article is to point out that the structural relation between the  $\eta$- and $\lm$-deformations is particularly explicit, obvious and neat in the framework of the theory of the  $\E$-models developed   in the context of the Poisson-Lie T-duality in \cite{KS96,KS97}.     In this regard, we wish 
to stress the conceptual and technical utility  of several  papers on the Poisson-Lie-T-duality like \cite{KS97,KS97b} which so far remain somewhat in the shadow of the initial works \cite{KS95,KS96}.  Indeed, as we shall show,  the results of the  paper \cite{KS97} permit to establish that not only the $\eta$-deformation but also their $\lambda$-counterpart belongs to the class of the $\E$-models introduced in \cite{KS96,KS97}.  In fact, the difference between $\eta$- and $\lambda$-deformations turns out to be given {\it solely} by  the choice  of the Drinfeld double  encoding the Hamiltonian structure of the integrable $\sigma$-model in question. The choice of the complexified group $G^\bc$  yields the $\eta$-deformation while the double $G\times G$ corresponds to the $\lambda$-deformation.

 The paper is organized as follows: In Section 4, we review the notion of the Drinfeld double current algebra as well as that of  the  $\E$-model \cite{KS96,KS97}.  In Section 5, we show that the  $\lambda$-model on arbitrary compact simple group target  $G$  is a particular case of the $\E$-model  and, for completeness, we review also the result of \cite{K02} establishing the same thing  for  the $\eta$-model. In Section 6,  we establish the result concerning the analytical continuation relation between the $\lm$ and the dual  $\eta$ target geometries for any $G$ and, finally, we devote Section 7 to a discussion of the results and to an outlook.
 
 \section{$\E$-models}
 
Let $\D$ denote a real finite dimensional Lie algebra  and let $(.,.)_\D$ be an ad-invariant non-degenerate symmetric bilinear form on $\D$. We then construct an infinite-dimensional Poisson manifold $P_\D$  the coordinates $j^A(\sigma)$ of which are labeled by one discret parameter $A=1,...,$dim$\D$ and  one continuous (loop) parameter $\sigma$, with the defining Poisson brackets given by 
\be \{j^A(\sigma), j^B(\sigma')\}=F^{AB}_{~~~C}j^C(\si)\delta(\si-\si')+D^{AB}\d_\si\delta(\si-\si').\label{dca}\ee
Here $F^{AB}_{~~~C}$ are the structure constants of $\D$ in some basis $T^A\in\D$ and
\be D^{AB}:=(T^A,T^B)_\D.\ee
The Poisson manifold $P_\D$  is referred to as the (symplectic\footnote{The invertibility of the Poisson tensor may fail  only  in a finite-dimensional zero mode sector in the
Fourier-transformed current components  $j^A(\sigma)$ which is determined by boundary conditions imposed on the currents.}  version of the)  current algebra associated to $\D$.
 
In what follows, we  shall study only quadratic Hamiltonians in  $j^A(\si)$  based on a choice of an $\br$-linear self-adjoint idempotent operator $\E:\D\to\D$ and given by the following formula
\be H_\E:=\jp\int d\si(j(\si),\E j(\si))_\D.\label{ham}\ee
Here we have used a   $\D$-valued coordinates $j(\si)$ on $P_\D$  defined by
\be (j(\si),T^A)_\D:=j^A(\si).\ee
We also state, for the completeness, that the self-adjointness and the idempotency of $\E$ (which are essential for the world-sheet Lorentz invariance of the Hamiltonian) mean, respectively
\be (\E x,y)_\D=(x,\E y)_\D, \quad \forall x,y\in\D; \qquad \E^2x=x, \quad \forall x\in\D.\ee 
{ \bf The dynamical system on the phase space $P_\D$ defined by the current algebra Poisson brackets
  \eqref{dca} and by the quadratic  Hamiltonian \eqref{ham} is referred to as an $\E$-model}. It was originally defined in 
\cite{KS96,KS97} and its equations of motion have the zero-curvature form valued in $\D$, that is
\be \d_\tau j=\d_\si(\E j)+[\E j,j].\label{zcd}\ee 
Here $\tau$ stands for the time.

\medskip

\noindent {\bf Remark 1}: {\small In \cite{KS96,KS97}, we have been using  a parametrization of the phase space $P_\D$ in terms of  a group-like variable $l(\si)$ taking values in the loop group of the Drinfeld double $D$. ($D$ is a Lie group the Lie algebra of which is $\D$.) The relation with the current algebra description $j(\si)$ reads
\be j(\si)=\d_\si l(\si)l(\si)^{-1}\ee
and the  equation of motion \eqref{zcd} takes form
\be \d_\tau ll^{-1}=\E\d_\si ll^{-1}.\ee
The Poisson brackets  expressed in terms of the variables $l(\si)$  are more cumbersome than the elegant current algebra formula \eqref{dca}, nevertheless, the
expression for the symplectic form on $P_\D$ is simpler in the $l(\si)$ language (see \cite{KS96,KS97} for details).}

\medskip

\noindent {\bf Remark 2}: {\small Suppose that there is a linear one-parameter family of the Lie algebra structures on the vector space $\D$, which means that the structure constants 
$F^{AB}_{~~~C}$ can be written as 
\be F^{AB}_{~~~C}= F^{AB}_{0~~C}+\e F^{AB}_{1~C}, \quad \e\in\br.\label{gen}\ee
Then the current algebra Poisson structure \eqref{dca} can be represented accordingly as
\be \{j^A(\sigma), j^B(\sigma')\}= \{j^A(\sigma), j^B(\sigma')\}_0 +\e \{j^A(\sigma), j^B(\sigma')\}_1.\ee
The  Poisson structures $\{.,.\}_0$ and $\{.,.\}_1$ appearing in this relation can be readily read off from  Eq. \eqref{dca} and they are
automatically compatible because the structure constants $F^{AB}_{~~~C}$ verify the Lie algebra Jacobi identity for every $\e$.}

\medskip 

Suppose now that   there is  a Lie subalgebra $\G\subset\D$  isotropic with respect to the bilinear form $(.,.)_\D$ and such that  dim$\G=\jp$dim$D$
(the isotropy means $(x,x)_\D=0, \forall x\in\G$). Then it was shown in \cite{KS97} that there is a non-linear $\sigma$-model on the target $D/G$ which  can be identified with  the 
$\E$-model $(P_\D,H_\E)$. Here  $G$ is the subgroup of $D$ corresponding to the subalgebra $\G$ and "can be identified" means the existence of a symplectomorphism (i.e. a canonical transformation) taking the phase space and the Hamiltonian of the $D/G$ $\si$-model onto $P_\D$ and
$H_\E$, respectively. The target space geometry of the $D/G$ model was worked out in detail in \cite{KS97,KS97b,K02} and it is encoded in the following action:
 \be S_\E(f)=S_{WZW,\D}(f) -\int d\xi^+ d\xi^-(P_f(\E) f^{-1}\d_+f,f^{-1}\d_-f)_\D.\label{emo}\ee
Here the action $S_{WZW,\D}(f)$ is given by
 $$  S_{WZW,\D}(f):=$$\be :=\jp\int  d\xi^+d\xi^-(f^{-1} \partial_+ f,f^{-1}\partial_-f)_\D+ \frac{1}{12}\int d^{-1}(dff^{-1},[dff^{-1},dff^{-1}])_\D,\ee
  the usual  light-cone variables $\xi^\pm$ and derivatives $\d_\pm$ read
\be \xi^\pm:=\jp(\tau\pm±\sigma), \qquad \d_\pm:=\d_\tau\pm\d_\sigma, \ee
 $f$ stands for the parametrization of the right coset $D/G$  by elements $f$ of $D$ (if there exists no global section of this fibration, we can choose several local sections covering the whole base space $D/G$) and, finally,
 $P_f(\E)$ is a projection from $\D$ into $\D$ defined by the relations
 $${\rm Im}P_f(\E)=\G,\qquad {\rm Ker}P_f(\E)=(\1+{\rm Ad}_{f^{-1}}\E{\rm Ad}_{f})\G. $$
 
 \medskip

\noindent {\bf Remark 3}: {\small
 The use of the projection $P_f(\E)$  in the formula \eqref{emo} is a new result (a by-line one) of  the present paper which encompasses the results of \cite{KS97,KS97b,K02} (e.g. the formula  (12) of \cite{KS97}) in a basis independent way.}
 
 \medskip

We do not repeat here the derivation of the formula  \eqref{emo} for the $\sigma$-model action from  the  $\E$-model data  $(P_\D,H_\E)$  as it is presented in \cite{KS97,KS97b,K02}  but  we do write down  the symplectomorphism associating to every solution of the equation of motion of the $\si$-model \eqref{emo}  the solution of the equation of motion \eqref{zcd} because this result is not contained in \cite{KS97,KS97b,K02}  :
\be j =\d_\si ff^{-1} -\jp f\left(P_f(\E)f^{-1}\d_+f-P_f(-\E)f^{-1}\d_-f\right)f^{-1}.\ee

\section{Current algebras of  $\eta$ and $\lm$  deformations}

Consider a simple compact real Lie algebra $\G$ equipped with its standard  Killing-Cartan form $(.,.)$. We introduce one-parameter family of real Lie-algebras $\D_\e$ which all have the property of being the Drinfeld doubles of $\G$. As the vector space, $\D_\e$ is just the direct sum of the vector space $\G$ with itself:
\be \D_\e:=\G\stc \G,\ee
the Lie algebra bracket $[.,.]_\e$ on $\D_\e$ is defined in terms of the commutator $[.,.]$ in $\G$ as follows
\be [x_1\stc x_2,y_1\stc y_2]_\e:=([x_1,y_1]+\e[x_2,y_2])\stc ([x_1,y_2] +[x_2,y_1]), \quad x_i,y_i\in\G,\label{aco}\ee
and, finally, the ad-invariant non-degenerate symmetric bilinear form  $(.,.)_\D$ does not depend on $\e$ and it is given by
\be (x_1\stc x_2,y_1\stc y_2)_\D:=(x_2,y_1)+(x_1,y_2).\label{bfd}\ee
Note that $\G$ is embedded in $\D_\e$ as $\G\stc 0$, or, said in other words, $\D_\e$ is the Drinfeld double of its subalgebra $\G\stc 0\simeq \G$. 

We now introduce a one-parameter family of  $\E$-models $(P_{\D_\e},H)$ based on the current algebra \eqref{dca} for the Drinfeld double $\D_\e$  and equipped with the quadratic Hamiltonian
\eqref{ham} given by the following  choice of the self-adjoint idempotent operator $\E$:
 \be \E(x_1\stc x_2):=(x_2\stc x_1). \label{dee}\ee
 Because here we speak about the particular operator $\E$ given by Eq. \eqref{dee}, we denote just by $H$ the  Hamiltonian  associated to it via \eqref{ham}, reserving the notation $H(\E)$  to  situations when a generic operator $\E$ occurs. 
 
 \medskip
 
 \noindent {\bf Remark 4}: {\small We  note  that the structure  constants of the Lie algebra $\D_\e$ have precisely the structure \eqref{gen} of Remark 2 which means that the symplectic structure of the $\E$-model $(P_{\D_\e},H)$ has  the form of the linear combination $\{.,.\}_0+\e\{.,.\}_1$  of two compatible Poisson structures as mentioned in the  Result 3 of the section Summary. }
 
 \medskip
 
 We now evaluate,  for every $\e$, the second order $\sigma$-model action \eqref{emo} of the $\E$-model $(P_{\D_\e},H)$.  We start with the simplest case $\e=0$ where it turns out to hold:
 
 \medskip

\noindent  {\it The  $\E$-model  $(P_{\D_0},H)$  can be identified with the principal chiral model on $G$.}

\medskip

Let us demonstrate this statement :

\medskip

 We first remark, that the Drinfeld double $D_0$  is
  the semi-direct product of manifolds $G$ and $\G$, i.e.  the  group law reads
 \be (g_1,x_1)(g_2,x_2)=(g_1g_2,x_1+g_1x_2g_1^{-1}), \qquad g_1,g_2\in G, \quad x_1,x_2\in\G.\label{bbl}\ee
It can be easily checked that, indeed, the law \eqref{bbl} gives rise to the Lie algebra commutator \eqref{aco} for $\e=0$.
 Now note that the commutation relation \eqref{aco}  implies 
 \be   [0\stc x_2,0\stc y_2]_0=0.\ee
 Denote the Abelian Lie algebra $0\stc\G$ by the symbol $\tilde\G$ and the corresponding Lie group  by $\tilde G$.  (The elements of $\tilde G$  are therefore $(e,x)\in D_0$, $e$ being the unit element of $G$.) 
 
 Consider now the $\si$-model \eqref{emo} on the target $D_0/\tilde G$. This coset can be obviously identified with the  subgroup $G$ of $D_0$, the elements of which are $f=(g,0)\in\D_0$.  Thus
 the field $f$ featuring in \eqref{emo} can be chosen to take values   $(g,0)\in D_0$. In this case the part $S_{WZW,\D}(f)$ of the action \eqref{emo} vanishes because the Lie algebra $\G$ of $G$ is maximally isotropic (i.e. $(\G\stc 0,\G\stc 0)_\D=0$).  Since the operator $\E$ given by \eqref{dee} evidently commutes with Ad$_{(g,0)}$, the projection $P_{(g,0)}(\E)$ does not depend on $g$ and it is easily found to be given by 
 \be P_{(g,0)}(\E)(x_1\stc x_2)=(0\stc (x_2-x_1)), \ee
 hence
 \be P_{(g,0)}f^{-1}\d_+f=P_{(g,0)}(g^{-1}\d_+g \stc 0) =(0\stc -g^{-1}\d_+g).\label{sof}\ee
 Combining \eqref{emo}, \eqref{bfd} and \eqref{sof} we find the following action of the $\sigma$-model  on $D_0/\tilde G$ :
 \be S_{\E,0} (g)= \int d\xi^+ d\xi^-(g^{-1}\d_+g,g^{-1}\d_-g).\ee
This is indeed the action of the principal chiral model on the group $G$ \cite{ZM}.

\medskip

Now we show that the evaluation of the second order $\sigma$-model action \eqref{emo} of  the  $\E$-models $(P_{\D_\E},H)$ for $\e>0$ gives the $\lm$-model of \cite{S14}.  More precisely, it holds

\medskip

\noindent {\it  For $\e>0$, the $\E$-model  $(P_{\D_\e},H)$  can be identified with the $\lm$-model on $G$  characterized by the action
$$ S_\lm(g)= \jp\int  d\xi^+d\xi^-(g^{-1} \partial_+ g,g^{-1}\partial_-g)+ \frac{1}{12}\int d^{-1}(dgg^{-1},[dgg^{-1},dgg^{-1}])$$\be +\lm\int d\xi^+d\xi^- (\d_+gg^{-1},(1-\lm{\rm Ad}_{g^{-1}})^{-1}g^{-1}\d_-g),\label{sfa}\ee
where \be \lm= \frac{1-\e^{\jp}}{ 1+\e^{\jp}}.\label{lm}\ee}

We start the argument by considering the Lie algebra $\G\oplus \G$ (i.e. the direct sum of the Lie algebra $\G$ with itself), the elements of which will be typically denoted $(\al_1,\al_2)$. There is an ad-invariant non-degenerate symmetric bilinear form on $\G\oplus \G$ given by the formula
\be ((\al_1,\al_2),\bt_1,\bt_2))_{\G\oplus \G}:= (\al_1,\bt_1)-(\al_2,\bt_2).\label{bgg}\ee
For each $\e$ positive there is an isomorphism of Lie algebras  $\Phi_\e: \D_\e \to \G\oplus \G$ given by 
\be \Phi_\e(x_1\stc x_2) =(x_1+\e^{\jp}x_2,x_1-\e^{\jp}x_2).\ee 
This isomorphism preserves the bilinear forms \eqref{bfd} and \eqref{bgg} up to normalization, that is 
\be(\Phi_\e(x),\Phi_\e(y))_{\G\oplus \G}= 2\e^{\jp}(x,y)_\D, \qquad x,y\in\D_\e.\ee
The existence of the isomorphism $\Phi_\e$ means that we can work   with the   double $\G\oplus \G$ instead of $\D_\e$, if we translate  by $\Phi_\e$ to the $\G\oplus \G$ context also the operator
$\E:\D_\e\to\D_\e$ given by \eqref{dee}. The translated operator $\E_\e:\G\oplus \G\to\G\oplus \G$ is  defined by the requirement
\be \E_\e\circ\Phi_\e=\Phi_\e\circ\E,\ee
which gives
\be \E_\e(\al,\bt)= \jp(\e^{\jp}+\e^{-\jp})(\al,-\bt) +\jp(\e^{\jp}-\e^{-\jp})(\bt,-\al).\ee
The group Drinfeld double of the Lie algebra $\G\oplus\G$ is evidently $G\times G$ (i.e.  the direct product of $G$ with itself) and  its elements will be typically denoted as $(a_1,a_2)$. The diagonal subgroup of $G\times G$ generated by the elements  of the form $(a,a)$ will be denoted as $G^\delta$. The corresponding Lie algebra $\G^\delta$ is maximally isotropic (it is the image of the subalgebra $\G\stc 0\subset \D_\e$ under the isomorphism $\Phi_\e$) and its elements are $(\al,\al)$. In order to apply to the present situation the  general formula \eqref{emo},  there remains to parametrize the cosets $D/G^\delta$ by the elements of $D$ and to identify the projection $P_f(\E_\e)$.  Obviously, the coset $D/G^\delta$ 
can be identified with the first copy $G$ in the direct product $G\times G$ which gives the parametrization $f=(g,e)$.  $P_{(g,e)}(\E_\e)$ is then straightforwardly found to be
equal to \be P_{(g,e)}(\E_\e)(\al,\bt)=(\frac{\lm}{\lm-{\rm Ad}_{g^{-1}}}\al+ \frac{1}{1-\lm{\rm Ad}_{g}}\bt, \frac{\lm}{\lm-{\rm Ad}_{g^{-1}}}\al+ \frac{1}{1-\lm{\rm Ad}_{g}}\bt), \label{pfe}\ee
where $\lm$ is given by the formula \eqref{lm}.

Finally, taking into account that $f^{-1}\d_+f=(g^{-1}\d_+ g,0)$, the wanted formula \eqref{sfa} follows directly (up to an overall normalisation) from Eqs. \eqref{emo}, \eqref{bgg} and \eqref{pfe}.
 
\medskip

\noindent {\bf Remark 5}: {\small Note that when the parameter $\e$ ranges from $0$ to $+\infty$, the parameter $\lm$ given by \eqref{lm} ranges from $-1$ to $1$. This is to be compared with the original paper \cite{S14} where the way of obtaining the action \eqref{sfa} (by a gauging procedure) leads to the interval of the values of $\lm$ between $0$ and $1$. Thus the vantage point based on the $\E$-models "sees" more possible values of $\lm$.}

\medskip

The fact that for $\e<0$  the evaluation of the second order $\sigma$-model action \eqref{emo} of  the  $\E$-models $(P_{\D_\E},H)$  gives the $\eta$-model of \cite{K02}  was proven already in \cite{K02}. However, to keep the exposition self-contained we outline here the argument :

 \medskip

Consider the Lie algebra $\G^\bc$ (i.e. the complexification of $\G$) the elements of which will be typically denoted as $z$. 
There is an ad-invariant non-degenerate symmetric bilinear form on $\G^\bc$ given by the formula
\be (z_1,z_2)_{\G^\bc}:= -\ri(z_1,z_2)+\ri\overline{(z_1,z_2)},\label{bgc}\ee
where $(.,.)$ is the Killing-Cartan form on $\G^\bc$ and $\overline{number}$ stands for the  complex conjugation of the $number$.

For each $\e$ negative, there is an isomorphism of Lie algebras  $\Psi_\e: \D_\e \to \G^\bc$ given by 
\be \Psi_\e(x_1\stc x_2) =x_1+\vert\e\vert^{\jp}{\ri} x_2.\ee 
This isomorphism relates the bilinear forms \eqref{bfd} and \eqref{bgc} up to normalization, that is 
\be(\Psi_\e(x),\Psi_\e(y))_{\G^\bc}= 2\vert\e\vert^{\jp}(x,y)_\D, \qquad x,y\in\D_\e.\ee
The existence of the isomorphisme $\Psi_\e$ means that we can work with the double $\G^\bc$ instead of $\D_\e$, if we translate to the $\G^\bc$ context also the operator
$\E:\D_\e\to\D_\e$ given by \eqref{dee}. The  translated operator $\E_\e:\G^\bc\to\G^\bc$ is   defined by the requirement
\be \E_\e\circ\Psi_\e=\Psi_\e\circ\E,\ee
which gives
\be \E_\e z= \frac{\ri}{2}(\vert\e\vert^{\jp}-\vert\e\vert^{-\jp})z-\frac{\ri}{2}(\vert\e\vert^{\jp}+\vert\e\vert^{-\jp})z^*.\label{sbl}\ee
Here $z^*$ stands for the Hermitian conjugation.

The group Drinfeld double of the Lie algebra $\G^\bc$ is evidently the complexified group $G^\bc$ viewed as the real group.  We shall evaluate the $\sigma$-model action \eqref{emo} on the target $G^\bc/\tilde G$ where for the $\tilde G$  we take the  isotropic  $AN$ subgroup of $G^\bc$  featuring in the standard Iwasawa decomposition $G^\bc\simeq GAN$ \cite{Zhel}.  It then follows that the space of cosets $G^\bc/\tilde G$ can be identified with the group $G$  thus the field $f$ in \eqref{emo} 
can be chosen $G$-valued: $f=g$. However, the operator $\E_\e$ as given by \eqref{sbl} obviously commutes with Ad$_g$ therefore the projection $\tilde P_{f=g}(\E_\e)$ does not depend
on $f$ (we put tilde  over $P(\E_\e)$ in order to indicate that the image of this projection is $\tilde\G$ and, in what follows, we suppress the subscript $f$). In order to find $\tilde P(\E_\e)$ explicitly, we  note that  the elements of $\tilde\G$ can be parametrized by the elements of $\G$ by using the so-called Yang-Baxter operator $R:\G\to\G$ (the explicit formula for $R$ can be found in \cite{K02,K09}).  Explicitly, every $\zeta\in\tilde\G$ can be uniquely written as
\be \zeta =(R-\ri)u\ee
for some $u\in\G$. With this insight, we find straightforwardly
\be \tilde P(\E_\e)z=\jp\frac{R-\ri}{1+\sqrt{\vert\e\vert}R}\left((\ri+\sqrt{\vert\e\vert})z+(\ri-\sqrt{\vert\e\vert})z^*\right).\ee
Taking into account the isotropy of the group $G$ (which eliminates the $S_{WZW,\D}(f)$ term from the action \eqref{emo}), applying $\tilde P(\E_\e)$ on $g^{-1}\d_+g$ and inserting the result  in the general formula \eqref{emo} we find 
\be S_\eta(g)= \jp\int  d\xi^+d\xi^-(g^{-1} \partial_+ g,(1-\eta R)^{-1}g^{-1}\partial_-g) ,\label{YB}\ee
where $\eta=\sqrt{\vert\e\vert}$. This coincides with the action of the $\eta$-model of Ref. \cite{K02,K09}.

We note finally, that in the present  Section 4 we have established the Results 2 and 3 as stated in the Section Summary.

\section{ T-duality and analytic continuation}
 
By   the Poisson-Lie T-dual of the $\eta$-model \eqref{YB} we shall mean the model \eqref{emo} based on the same $\E_\e$ operator \eqref{sbl} as the original model
\eqref{YB} but with the target space being $D/G$ instead of $D/\tilde G$. As in \cite{K02,K09}, we can identify the coset $D/G$ with the group $\tilde G=AN$ and, by setting $f=b\in AN$ and realizing that $S_{WZW,\D}(b)=0$, we trivially obtain from the basic formula  \eqref{emo} 
the  action  of the dual model in the following form 
\be \tilde S_\eta(b)=\jp\int  d\xi^+d\xi^-( \partial_+ bb^{-1},\tilde O(b)^{-1} \partial_-bb^{-1})_\D.\label{40}\ee
We do not specify further\footnote{The interested reader can find the explicit expression for $\tilde O(b)$  in \cite{K02} where $\tilde O(b)$ is related 
to the well-known Poisson-Lie structure $\tilde\Pi(b)$ on the group $AN$ via the formula $\tilde\Pi(b)=\tilde O(b)-\tilde O(\1)$.} the $b$-dependent  linear operator $\tilde O:\G\to\tilde\G$     because it is not the form \eqref{40} of the dual action that we are going to compare  with the $\lm$-model action \eqref{sfa}. Indeed, in trying to do so we would hurt on  a very complicated dependence of $\tilde O(b)$ on $b$.
Fortunately, we find in this paper a way out of these technical difficulties by  identifying the coset $D/G$ not with the group $AN$ but with the space $P$ of all  positive definite Hermitian
elements of the group $G^\bc$.  This new identification is based on the well-known  fact that 
 every element of $D=G^\bc$ admits a unique polar decomposition as the product of a  positive definite Hermitian element with an unitary element. From this statement it can be easily derived  that 
 the $AN$-parametrization and the $P$-parametrization of the coset $D/G$  is  related by the diffeomorphism  $\Upsilon:AN\to P$: \be\Upsilon(b)=\sqrt{bb^*}.\ee  
 
 To obtain the action of the dual model in the $P$-parametrization, it is now sufficient to set $f=\Upsilon(b)$  and to identify the projection $P_{\Upsilon(b)}(\E_\e)$:  \be P_{\ub}(\E_\e)z= \left(\sve -\ri +(\sve+\ri)\Ad_{bb^*}\right)^{-1}\left((\sve +\ri) Ad_{bb^*}z -  (\sve -\ri)z^*\right).\label{dp}\ee
Here $z^*$ means the Hermitian conjugation of the element $z$. 

Inserting \eqref{dp} and \eqref{bgc} into the basic  formula \eqref{emo} and taking   into account that $\ub$ is Hermitian (this gives e.g. $(\ub^{-1}\partial_+\ub,\ub^{-1}\partial_-\ub)_{\G^\bc}=0$) we obtain 
for the action of the dual $\eta$-model 
$$ \tilde S_\eta(b) =-2\ri S_{WZW}(\ub) +$$\be+2\ri\int  d\xi^+d\xi^-\left(\frac{\ri +(\eta+\ri)\Ad_{\ub}}{(\eta+\ri)\Ad_{\ub}+(\eta-\ri)\Ad_{\ub^{-1}}}\ub^{-1}\d_+\ub,\ub^{-1}\d_-\ub\right).\label{blq}\ee

Here $\eta=\sve$ and the action $S_{WZW}(\ub)$ apearing in \eqref{blq} is based on the ordinary Killing-Cartan form $(.,.)$ and not on $(.,.)_{\G^\bc}$. Explicitely,
$$ S_{WZW}(\ub):=\jp\int  d\xi^+d\xi^-(\ub^{-1} \partial_+ \ub,\ub^{-1}\partial_-\ub)+$$\be + \frac{1}{12}\int d^{-1}(d\ub\ub^{-1},[d\ub\ub^{-1},d\ub\ub^{-1}]).\label{pwz}\ee
Note that the hermiticity of $\ub$ implies that the dual action $\tilde S_\eta(b)$ is {\it real} inspite of the factor i standing in front of the r.h.s. of  \eqref{blq}. In particular, the WZW term in the r.h.s. of \eqref{pwz} is purely imaginary. Finally, we use  the Polyakov-Wiegmann formula \cite{PW}   
  \be S_{WZW}(bb^*)=2S_{WZW}(\ub)+\int  d\xi^+d\xi^-(\ub^{-1} \partial_- \ub, \partial_+\ub\ub^{-1}),\ee
  and the identity
  \be (bb^*)^{-1}\d_\pm (bb^*)= \ub^{-1}(\ub^{-1}\d_\pm \ub)\ub+\ub^{-1}\d_\pm \ub,\ee
  which gives together  
$$ \tilde S_\eta(b) =$$\be =-\ri S_{WZW}(bb^*) -\ri\lm\int d\xi^+d\xi^- ((1-\lm{\rm Ad}_{bb^*})^{-1}\d_+(bb^*)(bb^*)^{-1},(bb^*)^{-1}\d_-(bb^*))\label{ohn}\ee
  with \be \lm =\frac{1-\ri \eta}{1+\ri \eta}.\ee
 Comparing the resulting expression \eqref{ohn} with the $\lm$-model action $S_\lm$  given by the formula \eqref{1} or \eqref{sfa},
 we conclude
 \be \tilde S_\eta(b)=-\ri S_\lm(bb^*), \quad \lm =\frac{1-\ri \eta}{1+\ri \eta}.\ee
 
\medskip

Of course, the replacing the unitary argument  $g$ by the positive definite  Hermitian  argument $bb^*$ in the $\lm$-model action \eqref{1}  can be interpreted as a simple analytic continuation of the coordinates parametrizing the Cartan torus.
This is because both $g$ and $bb^*$ can be parametrized in the Cartan way:
\be g=hth^{-1}, \quad bb^*=hah^{-1},\ee
where $h$ is in $G$, $t$ is in the compact Cartan torus $T$ of $G$ and $a$ is in the noncompact part $A$ of the complex Cartan torus $T^\bc$ of $G^\bc$.
Note in this respect that here $A$  is the same $A$ which appears in the Iwasawa decomposition $G^\bc=GAN$.

As an example, let us  explicitely describe the analytic continuation from the non-compact to the compact Cartan torus  in the case of the group $SU(N)$ in which 
$A$  is formed by the  real diagonal matrices of the form 
\be a_{ij}=e^{\psi_i}\delta_{ij}, \quad \sum_j\psi_j=0, \quad i,j=1,\dots N.\ee
 The analytic continuation of the real Cartan coordinates $\psi_j$ to the strictly imaginary values i$\psi_j$  obviously transforms $a_{ij}$  into an element of the compact Cartan torus $T$ hence it switches from the positive definite Hermitian $bb^*$ to the unitary $g$.   
 
We note finally, that in the present  Section 5 we have established the Result 1 as stated in the Section Summary with the notation $p=bb^*$.
 
 \section{Conclusions and outlook}
 
We have identified the $\lm$-model on a simple compact Lie group $G$ as a particular case of the $\E$-model and we have used this result  to relate the $\lm$-model to the Poisson-Lie T-dual of the $\eta$-model by   the analytic continuation for any simple compact Lie group $G$.  We have also interpreted the $\lm$-model and the $\eta$-model as two branches of a single  one-parameter family of dynamical systems characterized by the same Hamiltonian but by  the varying Poisson brackets.  

It is probable that the framework of the $\E$-models will be useful to establish, for general $G$,  the analytic continuation relating  the two-parametric 
 $\lm$ models of Ref. \cite{SST} with the duals of the bi-Yang-Baxter models of Ref. \cite{K14}.   It is also plausible that  the dressing cosets generalization of the $\E$-models of Ref. \cite{KS96b}  will represent a suitable framework  for establishing 
the analytic continuation relation between  the $\eta$ and the  $\lm$ deformations of the   $\sigma$-models  living on   cosets of $G$. 
 
 \medskip 
 
  \vskip1pc
  
  \noindent {\bf Acknowledgement}: The author would like to thank the organizers  and the participants of the workshop "$\eta$ and $\lambda$ Deformations in Integrable 
Systems and Supergravity" at the ITP Bern, where some of the present results were presented prior to the publication, 
for  hospitality and for many  stimulating discussions.


\noindent 

\begin{thebibliography}{99}
 \bibitem{ABF}{G. Arutyunov, R. Borsato and S. Frolov, 
{\it $S$-matrix for strings on $\eta$-deformed $AdS_5 \times S^5$},  arXiv:1312.3542 [hep-th]}
\bibitem{ADT}{G. Arutyunov, M. de Leeuw and S. van Tongeren, 
{\it The exact spectrum and mirror duality of the $(AdS_5\times S^5)_\eta$ superstring}, Theor.Math.Phys. {\bf 182} (2015) 1, 23, arXiv:1403.6104 [hep-th]}
\bibitem{AM}{G. Arutyunov, D. Medina-Rincon, {\it  	
Deformed Neumann model from spinning strings on $(AdS_5\times S^5)_\eta$ }, JHEP {\bf 1410} (2014) 50, arXiv:1406:2536 [hep-th]}
\bibitem{BFHP}{J. Balog, P. Forg\'acs, Z. Horv\'ath and L. Palla, {\it A new family of $SU(2)$ symmetric  integrable $\sigma$-models}, Phys. Lett. {\bf B324} (1994) 403, 
hep-th/9307030}
 \bibitem{BBP}{A. Banerjee, S. Bhattacharya and K. Panigrahi, {\it 
Spiky strings in $\chi$-deformed $AdS$}, JHEP {\bf 1506} (2015) 057, arXiv:1503.07447 [hep-th]}
\bibitem{BKL}{V.V. Bazhanov, G.A. Kotousov and S.L. Lukyanov, {\it Winding vacuum energies in a deformed $O(4)$ sigma model}, Nucl. Phys. {\bf B889} (2014) 817, arXiv:1409.0449 [hep-th]}
 \bibitem{Ch}{I. V. Cherednik; {\it Relativistically invariant quasiclassical limits of integrable two-dimensional quantum models},
Theor. Math. Phys. {\bf 47} (1981) 422}
\bibitem{CMY}{P.M. Crichigno, T. Matsumoto and K. Yoshida, {\it Deformations of $T^{1,1}$  as Yang-Baxter sigma models}, JHEP {\bf 1412} (2014) 085,  arXiv:1406.2249 [hep-th]}
\bibitem{DMV13}{ F. Delduc, M. Magro and B. Vicedo, {\it  On classical q-deformations of integrable sigma-models},  JHEP{\bf 11} (2013) 192,
 arXiv:1308.3581 [hep-th]}
 \bibitem{DMV13b}{F. Delduc, M. Magro and  B. Vicedo, { \it  An integrable deformation of the $AdS_5 \times  S^5$  superstring action},
 arXiv:1309.5850 [hep-th]}
  \bibitem{DMV15}{F. Delduc, M. Magro and  B. Vicedo, {\it  Integrable double deformation of the principal chiral model}, Nucl. Phys. {\bf B891} (2015) 312-321, arXiv:1410.8066 [hep-th]}
  \bibitem{ER}{O.T. Engelung and R. Roiban, {\it On the asymptotic states and the quantum S matrix of the $\eta$-deformed $AdS_5 \times  S^5$ superstring}, JHEP {\bf 1503} (2015) 168, arXiv:1412.5256 [hep-th]}
  \bibitem{F}{V.A. Fateev, {\it The sigma model (dual) representation for a two-parameter family of integrable quantum field theories} Nucl. Phys. {\bf B473} (1996) 509}
  \bibitem{FOZ}{ V.A. Fateev, E. Onofri and A.B. Zamolodchikov, {\it The Sausage model (integrable deformations of $O(3)$ sigma model)}, Nucl. Phys. {\bf B406} (1993) 521}
 \bibitem{HMS}{T.J.Hollowood, J.L. Miramontes and D. Schmidtt, {\it An integrable deformation of the $AdS_5\times S^5$ superstring}, J. Phys. {\bf A47} (2014) 49, 495402, arXiv:1409.1538 [hep-th]}
\bibitem{H15}{B. Hoare, {\it Towards a two-parameter $q$-deformation of $AdS_3\times S^3\times M^4$ superstrings}, Nucl. Phys. {\bf B891} (2015) 259-295, arXiv:1411.1266 [hep-th]}
\bibitem{HRT}{
B.~Hoare, R.~Roiban and A.~A.~Tseytlin,
 {\it On deformations of $AdS_n\times S^n$ supercosets},
 JHEP {\bf 1406} (2014) 002,
 arXiv:1403.5517 [hep-th]}
\bibitem{HT}{B. Hoare and A. A. Tseytlin, {\it On integrable deformations of superstring sigma models related to} $AdS_n\times S^n$ {supercosets}, Nucl. Phys. {\bf B897} (2015) 448, arXiv:1504.07213[hep-th]}
\bibitem{ISST}{G. Itsios, K. Sfetsos, K. Siampos and A. Torrielli,{\it The classical Yang-Baxter equation and the associated Yangian symmetry of gauged WZW-type theories}, Nucl. Whys. {\bf B889} (2014) 64, arXiv:1409.0554 [hep-th]}
\bibitem{KMY}{T.Kameyama and K.Yoshida, {\it  	
Anisotropic Landau-Lifshitz sigma models from $q$-deformed $AdS_5\times S^5$ superstrings}, JHEP {\bf 1408} (2014) 110, arXiv:1405.4467 [hep-th]}
\bibitem{KY} {I. Kawaguchi and K. Yoshida,
 {\it Hybrid classical integrability in squashed sigma models}, Phys.Lett. {\bf B705} (2011) 251, arXiv:1107.3662  [hep-th]}
\bibitem{KMY14}{I. Kawaguchi, T. Matsumoto and K.Yoshida, {\it  Jordanian deformations of the $AdS_5\times S^5$  superstring}, JHEP {\bf 1406} (2014) 146,
arXiv:1401.4855 [hep-th]}
\bibitem{KK} {M. Khouchen and J. Kluso\v n, {\it Giant Magnon on Deformed  	
$AdS_3 \times  S^3$}, Phys. Rev. {\bf D90} (2014) 6, 066001, arXiv:1405.5017 [hep-th]}
\bibitem{KS95}{C. Klim\v c\'\i k and P. \v Severa, {\it Dual non-Abelian duality and the Drinfeld double},
Phys. Lett. {\bf B351}
(1995) 455, hep-th/9502122; C. Klim\v c\'\i k, {\it Poisson-Lie $T$-duality},
Nucl. Phys. (Proc. Suppl.) {\bf 
B46} (1996) 116, hep-th/9509095; P. \v Severa, 
{\it Minim\'alne plochy a dualita}, Diploma thesis, 1995, in Slovak} 
\bibitem{KS96}{C. Klim\v c\'\i k and P. \v Severa, {\it  Poisson-Lie T-duality and loop groups of Drinfeld doubles},  Phys. Lett. {\bf B372} (1996),  hep-th/9512040}
 \bibitem{KS97}{C. Klim\v c\'\i k and P. \v Severa, {\it Non-Abelian momentum-winding exchange}, Phys.Lett. {\bf B383} (1996) 281, hep-th/9605212}
 \bibitem{KS97b}{C. Klim\v c\'\i k and P. \v Severa, {\it Open strings and $D$-branes in WZNW model}, Nucl.Phys. {\bf B488} (1997) 653, hep-th/9609112}
\bibitem{KS96b}{C. Klim\v c\'\i k and P. \v Severa, {\it Dressing cosets}, Phys.Lett. B381 (1996) 56, hep-th/9602162}
  \bibitem{K02}{C. Klim\v c\'\i k, {\it Yang-Baxter $\si$-models
and dS/AdS T-duality}, JHEP {\bf 0212} (2002) 051,
hep-th/0210095}
\bibitem{K09}{C. Klim\v c\'\i k,  {\it On integrability of the Yang-Baxter $\sigma$-model}, J.Math.Phys. {\bf 50} (2009)  043508, arXiv:0802.3518 [hep-th]}
\bibitem{K14}{C. Klim\v c\'\i k, {\it Integrability of the bi-Yang-Baxter $\sigma$-model}, Lett. Math. Phys. {\bf 104} (2014) 1095, arXiv:1402.2105 [math-ph]}
 \bibitem{LRT}{O. Lunin, R. Roiban and A.A. Tseytlin, {\it 	
Supergravity backgrounds for deformations of $AdS_n\times S^n$  supercoset string models}, Nucl. Phys. {\bf B891} 106, arXiv:1411.1066 [hep-th]}
\bibitem{L}{S. L. Lukyanov, The integrable harmonic map problem versus Ricci flow, Nucl. Phys. {\bf B865} (2012) 308, arXiv:1205.3201  [hep-th]}
\bibitem{MY}{T. Matsumoto and K. Yoshida, {\it 	 Yang-Baxter deformations and string dualities}, JHEP 1503 (2015) 137, arXiv:1412.3658 [hep-th]}
\bibitem{MORSY}{T. Matsumoto, D. Orlando, S. Reffert, J. Sakamoto and K.Yoshida, {\it Yang-Baxter deformations of Minkowski spacetime}, arXiv:1505.04553 [hep-th]}
\bibitem{Moh}{N. Mohammedi, {\it On the geometry of classical integrable  two-dimensional nonlinear $\sigma$-models}, {\it Nucl.Phys.} {\bf B839} (2010) 420, arXiv:0806.0550 [hep-th]}   
\bibitem{PW}{A. Polyakov and P. Wiegmann, {\it Theory of nonabelian goldstone bosons in two dimensions}, Phys. Lett. {\bf B131} (1983) 121}
\bibitem{S14}{K. Sfetsos, {\it Integrable interpolations: From exact CFTs to non-Abelian T-duals}, Nucl. Phys. {\bf B880} (2014) 225, arXiv:1312.4560 [hep-th]}
\bibitem{SST}{K. Sfetsos, K. Siampos and D. Thompson, {\it Generalised integrable $\lm$- and $\eta$-deformations and their relation}, arXiv:1506.05784 [hep-th]}
\bibitem{S}{V. Suneeta, {\it The sausage sigma model revisited}, Class. Quant. Grav. {\bf 32} (2015) 11, 115005, arXiv:1409.4158 [hep-th]}
\bibitem{T}{ A. A. Tseytlin, {\it On A 'Universal' class of WZW type conformal models},
 Nucl.\ Phys.\ B {\bf 418}  (1994) 173,
 [hep-th/9311062]}
\bibitem{vT}{S. van Tongeren, {\it Yang-Baxter deformations, AdS/CFT, and twist-noncommutative gauge theory}, arXiv:1506.01023 [hep-th], {\it Integrability of the $AdS_5\times S^5$ superstring and its deformations}, J.Phys. {\bf A47} (2014) 433001, arXiv:1310.4854 [hep-th]}
\bibitem{V15}{B. Vicedo, {\it Deformed integrable $\sigma$--models, classical $R$-matrices and classical exchange algebra on Drinfel'd doubles}, J.Phys. {\bf A48} (2015) 35, 355203, arXiv:1504.06303}
\bibitem{ZM}{V.E. Zakharov and A.V. Mikhailov, {\it Relativistically invariant two-dimensional
model of field theory which is integrable by means of the inverse scattering method},
Sov. Phys. JETP {\bf 47} (1978) 1017}
 \bibitem{Zhel}{D.P. Zhelobenko, {\it Compact Lie  groups and their representations}, {\it Translations
 of Mathematical Monographs}  {\bf 40}, AMS, Providence, Rhode Island (1973)} 
\end{thebibliography}
\end{document}